\input phyzzx \def\e{\adveq\eqno{\rm (\chapterlabel.\the\equanumber)}}
\def\mysec#1{\equanumber=0\chapter{#1}}
\def\adveq{\global\advance\equanumber by 1}
\def\myeq{{\rm \chapterlabel.\the\equanumber}}
\def\rarrow{\rightarrow}

\def\semidirect{\mathrel{\raise0.04cm\hbox{${\scriptscriptstyle |\!}$
\hskip-0.175cm}\times}}


\def\ref#1{$^{[#1]}$}

\def\r#1{$[\rm#1]$}

\def\e{\adveq\eqno{\rm (\chapterlabel.\the\equanumber)}}
\def\mysec#1{\equanumber=0\chapter{#1}}
\def\adveq{\global\advance\equanumber by 1}
\def\myeq{{\rm \chapterlabel.\the\equanumber}}
\def\rarrow{\rightarrow}

\def\semidirect{\mathrel{\raise0.04cm\hbox{${\scriptscriptstyle |\!}$
\hskip-0.175cm}\times}}


\def\ref#1{$^{[#1]}$}

\def\r#1{$[\rm#1]$}

\overfullrule=0pt
\date{August, 2024}
\date{August, 2024}
\titlepage
\title{Equivalence of Deformations of Berglund H\"ubsch Mirror Pairs}
\author{Alexander A. Belavin$^{1}$ and Doron R. Gepner$^2$}
\vskip20pt
\line{\hfill $^1$ \it Landau Instiute for Theoretical Physics, Chernogolovka, Russia\hfill}
\line{\hfill $^2$ \it Department of Particle Physics, Weizmann Institute, Rehovot, Israel\hfill} 

\abstract
We investigate here the deformations of Berglund H\"ubsch loop and chain mirrors where the original
manifolds are defined in the same weighted projective space. We show that the deformations are 
equivalent by two methods. First, we map directly the two models to each other and show that the deformations are the same for $79$ "Good" models, but not for the $77$ "Bad" ones. We then 
investigate the orbifold of the mirror pair by the maximal symmetry group and show that the number of deformations is the same and that they are almost the same, i.e., the first four exponents of 
the deformations are identical. 
\endpage

\mysec{Introduction.}

Compactification of superstring from $10$ to $4$ dimensions, requiring  space time supersymmetry,
implies that the internal manifold is a complex manifold with vanishing first Chern class, so called
Calabi--Yau manifold \REF\CY{P. Candelas, G. Horowitz, A. Strominger and E. Witten, Nucl . Phys. B 258 (1985) 46.}\r\CY. This is the geometrical approach. A complementary  approach is the algebraic approach, where one builds the model around a solvable 
conformal field theory (CFT) \REF\Gep{D. Gepner, Nucl. Phys. B 296 (1988) 757.}\r\Gep. When the CFT is a tensor product of  $N=2$ minimal models, various geometries were identified, so called Fermat surfaces \REF\Quint{D. Gepner, Phys. Lett. B 199 (1987) 380.}\r\Quint. This was generalized to the geometry of 
weighted projective spaces in the works \REF\Mart{E. Martinec,  Phys. Lett. B 217 (1989) 431.}
\REF\Vafa{B. Greene, C. Vafa and N. Warner, Nucl. Phys. B 324 (1989) 371.}\r{\Mart,\Vafa}.

Another important observation is the mirror symmetry, where the same CFT corresponds to a different
geometry \REF\Mir{L. Dixon and D. Gepner, unpublished.} \REF\Ples{B.R.  Greene and M.R. Plesser, Nucl. Phys. B 338 (1990) 15.} \r{\Mir,\Ples}.
The Fermat geometry was generalized to  so called chain and loop models by Berglund and H\"ubsch (BH) 
\REF\BH{P. Berglund and T. H\"ubsch, Adv. Math. 9 (1998) 327.} 
\REF\Kraw{M. Krawitz, arXiv: 0906.0796 (2009).}\r{\BH,\Kraw}, where the mirror model is given by an orbifold of the	
transposed defining matrix.
The periods of the two models were shown to be the same in ref. \REF\BBK{A. Belavin, V. Belavin and
G. Koshevoy, Lett. Math. Phys. 111 (2021) 4, 93.}\r\BBK.

In this paper we investigate the equivalence or birationality of BH models where the loop and chain models are defined
on the same weighted projective space. At the level of the model itself this was already established
in ref. \REF\Bi{T.L. Kelly, Adv. Theor. Math. Phys. 17, no. 6,  (2013) 1425.} \REF\Sc{
M. Shoemaker, Comm. Math. Phys. 331, no. 2, (2014) 417.} \REF\Cl{
P. Clarke, Complex Manifolds 1 (2014) 45.}\r{\Bi,\Sc,\Cl}. 
For connection with Batyrev construction
\REF\Ba{V.V. Batyrev, J. Alg. Geom. 3 (1994) 493.}\r\Ba,
see \REF\Bat{A. Belavin and B. Eremin, arXiv 2010.07687
(2020).}\r\Bat.

We establish here this equivalence at the level of the deformations.
We do this in two ways. First, we map the loop and chain directly to each other. We see
for about half of the models, this leads to equivalence of the deformations. The other half is not
equivalent. The second method is building the orbifold of the maximal group at the level of the
mirror models. We see that in all the models the number of deformations is the same.

We study the dual groups of the models. Again, for half of the models very interesting groups emerge,
as duals of the minimal group. For the other half the dual group is always maximal.

\mysec{The loop and chain models.}

The manifold is defined by $W=0$, where,
$$W=\sum_{i=1}^5 \prod_{j=1}^5 x_j ^{M_{ij}},\e$$
and $M$ is an integer matrix of the three types,

\def\s{{s^\prime}}

Fermat:$ M_{ij}=\delta_{i,j} A_i$,

Loop: ${M_L}_{i,j}=s_i \delta_{i,j}+\delta_{i,j+1}$,
$\delta_{i,6}=\delta_{i,1}$,

Chain:   ${M_C}_{i,j}=\s_i \delta_{i,j}+\delta_{i,j+1}$,
$\delta_{5,6}=0$,

We concentrate here on the relation between the loop and chain models. Explicitly, the equations 
of these two types are,
$$W_L=x_1^{s_1} x_2+ x_2^{s_2} x_3+ x_3^{s_3} x_4+ x_4^{s_4} x_5+x_5^{s_5} x_1,\e$$
and
$$W_C=x_1^{\s_1} x_2+ x_2^{\s_2} x_3+ x_3^{\s_3} x_4+ x_4^{\s_4} x_5+x_5^{\s_5} ,\e$$

where $s$ and $\s$ are integers which obey the equation,
$$\sum_{j} M_{i,j} k_j=\sum_j k_j=d,\e$$
for both the loop and the chain manifolds, and $k_j$ are some integers. We assume that the same
weights, $k_i$, appears in both the loop and the chain models.

Eq. (2.4) also ensures that the chain or loop models are Calabi--Yau manifolds.
We also assume that
$M_{i,j}$ is such that 
$$k_5=1.\e$$
As we shall see, this is important to allow the equivalence of both the chain and loop models.
Then the equivalent chain model is given by
$$\s_i=s_i, {\ \rm for\ }i=1,2,3,4,   \qquad {\rm and \ }\s_5=s_5+k_1,\e$$

To get the list of the allowed manifolds we run over $s_i$ in some range to find all the solutions of
eq. (2.4), with $k_i$ integers and $k_5=1$. The range that we take is from $2$ to $40$
for $s_i$ and $i=1,2,3,4$ and $s_5$ from $2$ to $160$. This way we find $156$ solutions listed 
in section (3). This generalizes the $111$ solutions found in ref. \r\BBK.

The mirror manifold is obtained by taking the transpose of the matrix $M$. i.e.,
$$M_L\rarrow M_L^t,\qquad {\rm and \ } M_C\rarrow M_C^t.\e$$
we denote the transpose matrices by $M^{LT}$ and $M^{CT}$ for the loop and chain models,
respectively.

The weights of the mirror manifolds are denoted by $k^{(1)}_j$ and $k^{(2)}_j$ and are the
solutions of the equations
$$\sum_{j=1}^5 M_{i,j}^{LT} k^{(1)}_j=\sum_{j=1}^5  k^{(1)}_j,\e$$
for the loop model, for all $i$.
The same equation holds for the chain model with $M^{LT}\rarrow M^{CT}$ and
$k^{(1)}_j\rarrow k^{(2)}_j$. 

We concentrate here on the mirrors of the mirror chain and loop models described by the matrices 
$M^{LT}$ and $M^{CT}$.
The deformations of the chain and loop models correspond to the generations of the string theory,
whose number is $h_{2,1}$ in the cohomology. For the tranaposed loop, these are given by solutions of the equation,
$$\sum_{r=1}^5 m^h_r k^{(1)}_r=d^\prime=\sum_{r=1}^5 k^{(1)}_r,\e$$
and $m^h_r$ are integers which obey \REF\Kr{M. Kreuzer, Phys. Lett. B 328 (1994) 312.}\r\Kr,
$$0\leq m_r^h\leq s_r-1.\e$$
For the transposed chain, we have integers $m_r^{h\prime}$, which are solutions of the equation,
$$\sum_{r=1}^5 m^{h\prime}_r k^{(2)}_r=d^{\prime\prime}=\sum_{r=1}^5 k^{(2)}_r,\e$$
and obey 
$$0\leq m^{h\prime}_1\leq s_1^\prime-2\qquad m^{h\prime}_r\leq s_r^\prime-1 {\rm\ for\ } r\geq2,\e$$
or
$$m^{h\prime}_1=s^\prime_1-1,\quad m^{h\prime}_2=0,\quad m^{h\prime}_r\leq s_r^\prime-1,\ r\geq 3,\e$$
for the transpose chain model. 
Here, $h$ goes from $1$ to $h_{2,1}$, labeling the solutions, 
For the transposed loop model the deformations are given by
$$Z_h=\prod_{r=1}^5 x_r^{m_r^h}.\e$$
Similarly, for the transpose chain model the deformations are given by
are given by 
$$Z_h^\prime=\prod_{r=1}^5 x_r^{{m^\prime}^h_r}.\e$$

The same equations eq. (2.9--2.14), hold for the original chain and loop models with $d$ and $k$ instead
of $d^\prime$ and $k^\prime$. We see that the number of deformations of the original loop and
chain models are equal, which can be seen by explicit calculation.

\mysec{The direct map of the mirror pairs.}

To pass from the mirror loop to the mirror chain, we use the transformation,
$$Q=(M^{LT})^{-1} (M^{CT}),\e$$
where the matrix multiplication is assumed. The inverse transformation is given by
$$Q^\prime=(M^{CT})^{-1} (M^{LT}),\e$$
and the transformation is
$$x_j\rarrow \prod_{r=1}^5 y_r^{Q_{j,r}},\e$$
and the same with $Q^\prime$.

We apply the transformation, eq. (3.3), to the deformations of the loop or the chain models.
We find that the transformed deformations of the transposed chain model are given by
$$P_h=Z_h \bigg (x_j\rarrow \prod_{r=1}^5 y_r^{Q_{j,r}}\bigg),\e$$
and similarly from the transposed chain model to the transposed loop model, with
$Z_h^\prime$ and $Q^\prime$.

It turns out, that for some of the models, $P_h$ and $P_h^\prime$ are all integer exponents.
Further, for this models, $P_h$ and $P_h^\prime$ are exactly identical with the deformations
of the chain and loop models, respectively. We counted $79$ such models. For these models,
the transposed loop and chain models are completely isomorphic or birational. Namely, the first
four exponents are the same for all of the deformations of the mirror loop and mirror chain.
For the other models the transformed deformations are completely different.

The $79$ models are listed below, i.e., the values  of $s_i$ for these models. We call these models 
``Good" models.

\overfullrule=0pt
$$\displaylines{\{2, 2, 4, 18, 38\}, \{2, 2, 16, 6, 14\}, \{2, 2, 25, 5, 23\}, \{2, 3, 3, 
  30, 19\}, \{2, 3, 11, 4, 19\}, 
  \cr
  \{2, 4, 3, 14, 17\},
  \{2, 4, 10, 4, 8\}, \{2,
   4, 25, 3, 23\}, \{2, 5, 4, 6, 11\}, \{2, 6, 6, 6, 4\}, \{2, 6, 13, 3, 
  11\}, 
  \cr
  \{2, 9, 3, 12, 7\}, \{2, 10, 4, 4, 20\}, \{2, 12, 3, 8, 9\}, \{2, 12, 
  9, 3, 7\}, \{2, 18, 3, 6, 13\}, \{2, 30, 3, 5, 21\}, 
  \cr
  \{2, 32, 5, 3, 
  19\}, \{3, 3, 2, 28, 21\}, \{3, 3, 9, 3, 10\}, \{3, 7, 2, 12, 9\}, \{3, 11, 
  2, 6, 13\}, \{3, 23, 2, 4, 25\}, 
  \cr
  \{3, 27, 4, 2, 21\}, \{4, 2, 2, 16, 
  40\}, \{4, 2, 3, 6, 21\}, \{4, 2, 6, 6, 6\}, \{4, 2, 21, 3, 27\}, \{4, 8, 2,
   4, 10\}, 
   \cr
    \{4, 14, 3, 2, 13\}, \{5, 3, 5, 3, 6\}, \{5, 23, 2, 2, 25\}, \{6, 
  2, 2, 8, 22\}, \{6, 3, 9, 2, 31\}, \{6, 4, 2, 6, 6\},
  \cr
  \{6, 4, 4, 2, 
  20\}, \{6, 6, 4, 2, 6\}, \{6, 14, 2, 2, 16\}, \{8, 3, 6, 2, 21\}, \{11, 3, 
  4, 2, 29\}, \{2, 2, 4, 14, 146\}, 
  \cr
  \{2, 2, 5, 9, 115\}, \{2, 2, 8, 6, 
  118\}, \{2, 2, 13, 5, 155\}, \{2, 3, 3, 14, 147\}, \{2, 3, 8, 4, 97\}, \{2, 
  4, 3, 9, 117\}, 
 \cr
  \{2, 4, 15, 3, 153\}, \{2, 7, 3, 6, 123\}, \{2, 7, 4, 4, 
  101\}, 
  \cr
  \{2, 12, 3, 5, 165\}, \{2, 14, 5, 3, 163\}, \{3, 3, 2, 12, 
  149\}, \{3, 3, 5, 3, 74\}, 
  \cr
  \{3, 5, 2, 6, 115\}, \{3, 5, 3, 3, 74\}, \{3, 5, 
  14, 2, 163\}, \{3, 11, 2, 4, 157\}, \{3, 15, 4, 2, 153\}, 
  \cr
  \{4, 2, 2, 12, 
  148\}, \{4, 2, 3, 5, 85\}, \{4, 2, 4, 4, 86\}, \{4, 2, 11, 3, 157\}, \{4, 4,
   2, 4, 86\}, 
   \cr
   \{4, 4, 7, 2, 101\}, \{4, 8, 3, 2, 97\}, \{5, 3, 2, 4, 
  85\}, \{5, 3, 12, 2, 165\}, \{5, 13, 2, 2, 155\},
  \cr
   \{6, 2, 2, 6, 112\}, \{6, 
  2, 5, 3, 115\}, \{6, 3, 7, 2, 123\}, \{6, 8, 2, 2, 118\}, \{9, 3, 4, 2, 
  117\}, 
  \cr
  \{9, 5, 2, 2, 115\}, \{12, 2, 2, 4, 148\}, \{12, 2, 3, 3, 
  149\}, \{14, 3, 3, 2, 147\}, \{14, 4, 2, 2, 146\}
  }
$$

The other models, which are not birational, are listed below, i.e. their $s_i$ values. There are $77$ such models. We call these models ``Bad" models.

$$\displaylines{
\{2, 2, 4, 28, 20\}, \{2, 2, 5, 13, 19\}, \{2, 2, 10, 10, 8\}, \{2, 2, 13, 
 7, 11\}, \{2, 3, 5, 10, 7\}, 
 \cr
 \{2, 5, 5, 4, 31\}, \{2, 7, 4, 8, 5\}, \{2, 8, 
 4, 4, 38\}, \{2, 8, 7, 4, 5\}, \{2, 11, 5, 4, 7\},
 \cr
  \{2, 17, 4, 4, 11\}, \{3,
  2, 3, 9, 29\}, \{3, 2, 3, 11, 18\}, \{3, 2, 5, 5, 25\}, \{3, 2, 9, 5, 
 9\}, 
 \cr
 \{3, 2, 13, 4, 14\}, \{3, 4, 3, 5, 8\}, \{3, 5, 3, 6, 5\}, \{3, 5, 21, 
 2, 30\}, \{3, 6, 3, 3, 22\}, 
 \cr
 \{3, 6, 5, 3, 5\}, \{3, 6, 13, 2, 18\}, \{3, 8,
  9, 2, 12\}, \{3, 10, 3, 3, 9\}, \{3, 11, 5, 2, 34\},
 \cr
  \{3, 12, 7, 2, 
 9\}, \{3, 13, 5, 2, 20\}, \{3, 17, 5, 2, 13\}, \{4, 2, 2, 26, 22\}, \{4, 4, 
 4, 4, 4\}, 
 \cr
 \{4, 4, 8, 2, 38\}, \{4, 4, 11, 2, 17\}, \{4, 5, 2, 8, 7\}, \{4, 
 5, 5, 2, 31\}, \{4, 8, 5, 2, 7\}, 
 \cr
 \{5, 2, 7, 4, 8\}, \{5, 2, 13, 3, 
 17\}, \{5, 3, 19, 2, 32\}, \{5, 4, 7, 2, 11\}, \{5, 6, 3, 2, 24\}, 
 \cr
\{5, 7, 
 3, 2, 14\}, \{5, 9, 3, 2, 9\}, \{7, 2, 3, 5, 10\}, \{7, 2, 5, 3, 34\}, \{7, 
 2, 9, 3, 12\}, 
 \cr
 \{7, 4, 5, 2, 8\}, \{7, 11, 2, 2, 13\}, \{8, 2, 2, 10, 
 10\}, \{9, 3, 7, 2, 12\}, \{10, 8, 2, 2, 10\}, 
 \cr
 \{19, 2, 3, 3, 30\}, \{20, 2,
  2, 4, 28\}, \{21, 3, 3, 2, 28\}, \{22, 4, 2, 2, 26\}, \{2, 2, 4, 16, 
 56\}, 
 \cr
 \{2, 2, 5, 10, 43\}, \{2, 2, 7, 7, 41\}, \{3, 2, 3, 8, 51\}, \{3, 2, 
 7, 4, 53\}, \{3, 5, 15, 2, 87\}, 
 \cr
 \{3, 5, 17, 2, 49\}, \{3, 6, 9, 2, 
 64\}, \{3, 7, 7, 2, 59\}, \{3, 10, 5, 2, 62\}, \{4, 2, 2, 14, 58\}, 
 \cr
 \{5, 2, 
 7, 3, 65\}, \{5, 3, 13, 2, 89\}, \{5, 3, 15, 2, 51\}, \{7, 4, 3, 2, 
 47\}, \{7, 7, 2, 2, 41\}, 
 \cr
 \{10, 5, 2, 2, 43\}, \{13, 2, 3, 3, 81\}, \{14, 2,
  2, 4, 58\}, \{15, 2, 3, 3, 47\}, \{15, 3, 3, 2, 79\}, 
  \cr
  \{16, 4, 2, 2, 
 56\}, \{17, 3, 3, 2, 45\}.
 \cr
 }$$
 
 Let us give an example for the biratonality of the loop and chain models. We take for the exponents,
$$s_i=\{2, 6, 6, 6, 4\}.\e$$
This is a Good model. 
We have here $k=\{3,1,1,1,1\}$, $k^{(1)}=\{97,25,37,35,53\}$ and $k^{(2)}=\{216,36,66,61,53\}$.
There are exactly two deformations of the transposed loop which are given by,
$$\{x[2]^2 x[3]^2 x[4]^2 x[5], x[1] x[2] x[3] x[4] x[5]\}.\e$$
There are also exactly two deformations of the transposed chain model,
$$\{y[2]^2 y[3]^2 y[4]^2 y[5]^2, y[1] y[2] y[3] y[4] y[5]\}.\e$$
We see, indeed, that the first four exponents of the deformations are the same for the two models.
This is general for all Good models.

The matrix which takes us from the loop to the chain model is given by, eq. (3.1),
$$Q=\pmatrix{1 & 0 & 0 & 0&  -216/247\cr 0 &  1 &  0 & 0 &  36/247\cr 0 &  0 & 1 & 
  0 &  -6/247 \cr  0 &  0 & 0 &  1 & 1/247\cr 0 & 0 & 0 & 0 & 432/247 \cr}.\e$$
 The transformation which takes us from the chain to the loop model is given by, eq. (3.4),
 $$\eqalign {& \{x[1] \rarrow  y[1]/y[5]^{216/247}, x[2] \rarrow  y[2] y[5]^{36/247}, 
 x[3] \rarrow  y[3]/y[5]^{6/247}, \cr  & \qquad  x[4] \rarrow y[4] y[5]^{1/247}, 
 x[5]  \rarrow  y[5]^{432/247}\}}\e$$
 
 Implementing this transformation to the transpose loop deformations, we find exactly the transposed
 chain deformations.
 
 Going the other way, from the transposed chain model to the transposed loop model, eq. (3.2), we find the matrix $Q^\prime$, which is 
 $$Q^\prime=\pmatrix{1 & 0 & 0 & 0 & 1/2 \cr 0 & 1 & 0 & 0 & -1/12\cr 0 & 0 & 1 &  0 & 1/72 \cr 0 & 0 & 
  0 & 1 & -1/432 \cr 0 & 0 & 0 & 0 & 247/432\cr}. \e$$
 The transformation from the chain to the loop models is given by, eq. (3.4),
$$\eqalign {& x[1] \rarrow  y[1] \sqrt{y[5]}, x[2] \rarrow y[2]/y[5]^{1/12}, 
 x[3] \rarrow  y[3] y[5]^{1/72},\cr & \qquad  x[4] \rarrow  y[4]/y[5]^{1/432}, 
 x[5] \rarrow  y[5]^{247/432}.}\e$$

Again, we find that the deformations of the transposed chain are mapped precisely to the deformations of the transposed loop. 

\mysec{The equivalence of the mirror pairs deformations under an orbifold of the maximal group.}

 Let us describe now the automorphism group of the mirror models. Denote, as before, eq. (2.7),
 the matrices $M^{CT}$ and $M^{LT}$ the defining matrices of the mirror chain and mirror loop.
 Denote the matrices by
 $$B={M^{LT}}^{-1},\e$$
 and
 $$B^\prime={M^{CT}}^{-1}.\e$$
Then the automorphism group of the mirror loop is generated by $q_i$,
$$q_i(X_j)=\exp(2\pi I B_{ji}) X_j,\e$$
and for the mirror chain,
$$p_i(X_j)=\exp(2\pi I B_{ji}^\prime) X_j.\e$$
This includes all the phase changes which preserve the super potential for the mirror chain and loop.

The elements of the automorphism group for the mirror loop are given by,
$q_i^{m_i}$, where $m_i$ are any non negative integers. Similarly for the mirror chain,
$p_i^{m_i}$.

We would like to ensure, in both cases, that the perturbation
$$\prod_{i=1}^5 X_i,\e$$
is invariant under the automorphism, since it is proportional to the $(3,0)$ form which is
needed for a Calabi Yau manifold. This implies that
$$\prod_{j=1}^5 q_i(X_j)=\prod_{j=1}^5 \exp(2 pi I B_{ji} m_i)\prod_{j=1}^5 X_j=\exp(2\pi I \sum_{i=1}^5 m_i k_i/d)\prod_{j=1}^5 X_j,\e$$
where we used the fact that
$$\sum_j B_{ji}=k_i/d.\e$$
Thus to ensure the invariance we require that
$$\sum_{i=1}^5 m_i k_i/d\in Z,\e$$
is an integer. 
Similar equation holds for the mirror chain model with $p$ instead of $q$. We denoted $I=\sqrt{-1}$.

We wish to investigate now, which of the mirror loop and chain deformations are invariant
under the automorphism group, since mirror holds only after modding out by this automorphism group.
Denoting as before the mirror loop deformations by $Z_h$ and $Z_h^\prime$, eqs. (2.12, 2.15),
we find that the surviving deformations for the mirror loop are given by $m^h_r$, which obey
$$  \sum_{i,j=1}^5  m^h_i B_{i j} m_j\in Z,\e$$
for any $m_j$ that solves eq. (4.8).
Similar equation holds for the mirror chain model with $m^h_i$ replaced by ${m^h}^\prime_i$.
We denote the solutions of eq. (4.9) by $n_i^h$ for the loop and ${n_i^h}^\prime$ for the chain.

We find that indeed the invariant deformations of the mirror loop and the mirror chain are of the same
number for all $156$ theories, including both the Good and Bad models. This establish the birationality at the level of the deformations.

Actually, there is a difference here between the Good and Bad models. For the Good models, all the 
deformations obey eq. (4.9) and thus,
$n_i^h=m_i^h$, for the loop, and similar for the chain, ${n_i^h}^\prime={m_i^h}^\prime$.
For the Bad models, this equality no longer holds and either $n_i^h$ or ${n_i^h}^\prime$ are no
longer the same as the deformations, $m_i^h$ or ${m_i^h}^\prime$.

One can see this also on the level of the dual symmetry group $\bar G$, which is defined by \r{\Kraw,\Bat},
$$\bar G=\left\{\prod_{i=1}^5 {(q^t_i)}^{r_i} \big| [r_1,\ldots, r_5] B[m_1,m_2,\ldots,m_5]^t \in Z{\rm\   for \ all\ } \prod_{i=1}^5 q_i^{m_i}\in G\right\},\e$$
Here $q_i$ and $q_i^t$ are exponents of  the rows and columns of $B$, respectively, eq. (4.3). $G$ is the group generated by $q_i^{m_i}$, eq. (4.8). Analogously we define $\bar G$ for the chain models, with 
$p_i$ replacing $q_i$, eq. (4.4).

We have two possibilities for the group $G$. The maximum one is the group generated by
$m_i$ where 
$$m_i k_i /d \in Z,\e$$
and the group is all the elements $q_i^{m_i}$. We denote this group as $G^{\rm max}$. 
The minimal group is the group generated by $\prod_i X_i$, which is always present.
This group is given by
$$G^{\rm min}=q_i^{\{w,w,w,w,w\}},\qquad{\rm where\ } w\in Z.\e$$

Now we can calculate the dual groups in both cases.  For the group $H={\bar G}^{\rm max}$ we find
the minimal dual group which is given by
$$H=\{Exp[2\pi I k_i w /d]\},\quad w=0,1,\ldots, d-1,\qquad i=1,2,3,4,5,\e$$
which is an order $d$ cyclical group.
This is the group $H$ for both the Good and Bad models.

For the minimal group we find the dual to be $Q$ where $H\subset Q$. For the Good models we find
that $Q=H$, i.e., the minimal and maximal dual are the same.
However, for the bad model, the group $Q$ is larger than $H$, either for the mirror loop or the mirror chain. It is always a cyclical group with order $z$ where $d | z$, z is a multiple of $d$.
We give an example of a dual Bad group below.

Let us give now an example for the birationality of the orbifold. The example we give is defined
by its exponents,
$$s_i=\{2,4,10,4,8\}.\e$$
In this case the deformations of the mirror loop and mirror chain are the same before and after the
orbifoldation, which we observe to hold for all the Good theories. The deformations are given here, for the loop,
$$ \{0,0,6,0,5\},\{0,1,4,1,3\},\{0,2,2,2,1\},\{1,0,3,0,3\},\{1,1,1,1,1\},\e$$
and for the chain,
$$\{0,0,6,0,8\},\{0,1,4,1,5\},\{0,2,2,2,2\},\{1,0,3,0,4\},\{1,1,1,1,1\}.\e$$
Here, the exponents are denoted as $\{m^h_i,\,i=1,\ldots,5\}$. 
Note that the chain and loop orbifoldized deformations are the same, except in the fifth number, which
may be different. This observation is general for all the theories. However, we do not have a proof for this.

Let us give an example of a Bad theory. We take
$$s_i=\{2, 3, 5, 10, 7\}.\e$$
Here, before orbifoldation, we have $4$ deformations for the mirror loop and $126$ for the mirror chain.
After orbifoldation there are $4$ of each, in agreement with the birationality. These are given by,
for the loop,
$$\{0,0,3,3,2\},\{0,1,0,5,3\},\{0,2,2,2,1\},\{1,1,1,1,1\},\e$$
and for the chain,
$$\{0,0,3,3,3\},\{0,1,0,5,5\},\{0,2,2,2,2\},\{1,1,1,1,1\}.\e$$
Again, the exponents are the same except possibly for the fifth one. For the Bad theories, we see that
the number of deformations before and after orbifoldation are not the same. However, they are the same after the orbifoldation.

The dual group here of the minimal group, $Q$, is a cyclical group of order $275$ for the mirror chain
and the maximal group for the mirror loop, of order $d=11$. 
The group $Q$ is generated by (one of the generators),
$$r_i=\{-2/275, 4/275, -12/275, 12/55, 9/11\}.\e$$
It would be of interest to study these groups.

\mysec{Discussion.}
In this paper we discussed the phenomena of loop and chain models defined on the same weighted 
projective space. We discussed  the birationality of the mirror loop and mirror chain deformations  in three different ways. First, we mapped the models directly on each other. We find that for the $79$ ``Good"
models the deformations map directly on each other. For   the $77$ ``Bad" models they do not map
and the mapping gives a bad result, namely, fractional and negative exponents.

The second way was to construct the orbifold under the maximal group of both the Good and Bad
models. We find that the deformations always map to each other, where the first four exponents are
the same and the fifth may be different. For the Good models the orbifold is identical to the 
untwisted deformations, which is not true for the Bad models.

Our final approach involved computing the dual group for both the maximal and minimal groups. This analysis yielded an interesting result for Bad models: the dual of the minimal group was larger than the minimal dual.
However the dual groups remained minimal for  maximal or  minimal groups in the original  Good theories,
because these groups coincide  in the Good cases",

\ack
A. Belavin thanks  the Weizmann institute for the hospitality in the visiting faculty program 
while part of this work was done. We thank I. Deichaite for remarks on the manuscript.

\refout

\bye